\documentclass{article}
\begin{document}
\begin{large}
\pagenumbering{arabic}

\begin{center}
\bigskip
{\large\bf Deep inelastic $J/\Psi$ production at HERA\\
in the colour singlet model with $k_T$-factorization}\\

\vspace{2cm}

\renewcommand{\thefootnote}{1)}
{\large 
A.V.~Lipatov \footnote{E-mail: artem\_lipatov@mail.ru}\,

{\it Physical Department, M.V. Lomonosov Moscow State University,\\
119992 Moscow, Russia\/}\\[3mm]}

\vskip 0.5cm

\renewcommand{\thefootnote}{2)}
N.P.~Zotov \footnote{E-mail: zotov@theory.sinp.msu.ru}

{\it D.V.~Skobeltsyn Institute of Nuclear Physics,\\ 
M.V. Lomonosov Moscow State University,\\
119992 Moscow, Russia\/}\\[3mm]

\end{center}

\vspace{1.5cm}

\begin{center}
{\bf{Abstract}}
\end{center}
\bigskip

We consider $J/\Psi$ meson production in $ep$ deep inelastic scattering
in the colour singlet model using the $k_T$-factorization QCD approach.
We investigate the $z$-, $Q^2$-, ${\bf p}_T^2$-, $y^*$- and $W$-dependences of
inelastic $J/\Psi$ production on different forms of the unintegrated 
gluon distribution. The ${\bf p}_T^2$- and $Q^2$-dependences of the 
$J/\Psi$ spin aligment parameter $\alpha$ are presented also. 
We compare the theoretical results with recent experimental data taken by 
the H1 and ZEUS collaborations at HERA. It is shown that experimental 
study of the polarization $J/\Psi$ mesons at $Q^2 < 1\,{\rm GeV}^2$ is 
an additional test of BFKL gluon dynamics.

\newpage

\section{Introduction} \indent 

It is known that from heavy quark and quarkonium production processes one 
can obtain unique information on gluon structure function of the proton 
because of the dominance of the photon-gluon or gluon-gluon fusion 
subprocess in the framework of QCD~[1--4]. Studying of  gluon distributions 
at modern collider energy (such as HERA, Tevatron) is important for 
prediction of heavy quark and quarkonium production cross sections at 
future colliders (LHC, THERA). At the energies of HERA 
and LEP/LHC colliders heavy quark and quarkonium production processes are 
so called semihard processes~[5]. In such processes by definition the hard 
scattering scale $\mu \sim m_Q$ is large compare to the $\Lambda_{{\rm QCD}}$ 
parameter but on the other hand $\mu$ is much less than the total 
center-of-mass energy: $\Lambda_{{\rm QCD}}\ll \mu\ll\sqrt s$. The last 
condition implies that the processes occur in small $x$ region: 
$x\simeq m_Q/\sqrt s\ll 1$, and that the cross sections of heavy quark and 
quarkonium production processes are determined by the behavior of gluon 
distributions in the small $x$ region.

It is also known that in the small $x$ region the standard parton model 
(SPM)
assumptions about factorization of gluon distribution functions and 
subprocess cross sections are broken because the subprocess cross 
sections and gluon structure functions depend on a gluon transverse 
momentum $k_T$~[6--8]. So calculations of heavy quark production cross 
sections 
at HERA, Tevatron, LHC and other collider conditions are necessary 
to carry out in the so called $k_T$-factorization (or semihard) QCD approach, 
which is more preferable for small $È$ region than SPM.

The $k_T$-factorization QCD approach is based on Balitsky, Fadin, Kuraev, Lipatov 
(BFKL)~[9] evolution equations. 
The resummation of the terms $\alpha_{S}^n\,\ln^n(\mu^2/\Lambda_{{\rm QCD}}^2)$,
$\alpha_{S}^n\,\ln^n(\mu^2/\Lambda_{{\rm QCD}}^2)\,\ln^n(1/x)$ and $\alpha_{S}^n\,\ln^n(1/x)$ 
in the
$k_T$-factorization approach leads to the unintegrated (depends from 
${\bf q}_T$) gluon distribution $\Phi(x,{\bf q}_T^2,\mu^2)$ which 
determine the probability to find a gluon carrying the longitudinal momentum 
fraction $x$ and transverse momentum ${\bf q}_T$ at probing scale $\mu^2$.
The unintegrated gluon distribution reduce to the conventional gluon 
distribution $xG(x,\mu^2)$ once the ${\bf q}_T$ dependence is integrated out:
\begin{equation}
xG(x,\mu^2) = xG(x,Q_0^2) + \int\limits^{\mu^2}_{Q_0^2} \Phi(x,{\bf q}_T^2,\mu^2)\, d{{\bf q}_T^2}.
\end{equation}

To calculate the cross section of a physical process the unintegrated gluon
distributions have to be convoluted with off mass shell matrix elements
corresponding to the relevant partonic subprocesses~[6--8]. In the off mass
shell matrix element the virtual gluon polarization tensor is taken in
the BFKL form~[5]:
\begin{equation}
L^{\mu\,\nu}(q) = {q_T^{\mu}\,q_T^{\nu}\over {\bf q}_T^2}.
\end{equation}

Nowadays, the significance of the $k_T$-factorization QCD approach
becomes
more and more commonly recognized~[10]. It is already was used for the 
description 
of a wide class heavy quark and quarkonium production processes~[6, 11--25], 
in particular, for the description heavy quark~[6, 11, 14, 16, 20--23]
and quarkonium~[13, 15, 19, 24, 25] photo and leptoproduction. 
It is notable that calculations in $k_T$-factorization QCD approach results 
to some effects which are absent in other approaches. It is a more fast 
growth of total cross sections in comparison with SPM, a broadering 
of the $p_T$ spectra due to extra transverse momentum of the colliding 
partons and other polarization properties of final particles in comparison 
with SPM.

We point out that heavy quark and quarkonium cross section calculations 
within the SPM in the fixed order of pQCD have some problems.
For example, the very large discrepancy (by more than an order of 
magnitude)~[26, 27] between the pQCD predictions for hadroproduction $J/\Psi$   
and $\Upsilon$ mesons and experimental data at Tevatron was found.
This fact has resulted in intensive theoretical investigations of such 
processes. In particular, it was required to enter additional transition 
mechanism from $c\bar c$-pair to the $J/\Psi$ mesons, so-called the colour 
octet 
(CO) model~[28], where $c\bar c$-pair is produced in the color octet state and 
transforms into final colour singlet (CS) state by help soft gluon radiation. 
The CO model was supposed to be applicable to heavy quarkonium hadro and 
electroproduction processes. However, the contributions from the CO 
mechanism to the $J/\Psi$ meson photoproduction contradict the 
experimental data at HERA for $z$-distribution~[29--32].

Another difficulty of the CO model is $J/\Psi$ polarization properties 
in $p\bar p$-interactions at Tevatron. In the framework of 
the CO model $J/\Psi$ mesons should be transverse polarized at the 
large 
transverse momenta ${\bf p}_T$. However, this fact contradicts the experimental 
data 
too.

The CO model has been applied earlier in analysis of the $J/\Psi$ 
inelastic production experimental data at HERA~[33, 34]. However, this
results are not agree with each other~[34]. It is notable that results 
obtained within the usual collinear approach and CS model~[35--38] 
underestimate experimental data by factor about 2.

Recently the first attempts to investigate a $J/\Psi$ polarization problem
in $p\bar p$-interactions at Tevatron in $k_T$-factorization approach 
were made~[24, 39, 40]. Also the theoretical prediction within semihard QCD 
approach~[19] are stimulated the experimental analysis $J/\Psi$ polarization 
properties at HERA conditions. However the further theoretical 
and experimental studies of this problem are necessary.

Based on the above mentioned results we consider here 
the deep inelastic $J/\Psi$ production at HERA
in the framework of the CS model and the $k_T$-factorization approach 
with emphasis of a role of a proton gluon distribution 
function. Using the CS model and 
formalism of the projection operator~[41], we investigate the 
$Q^2$-, ${\bf p}_T^2$-, $z$-, $y^*$- and $W$-dependences of $J/\Psi$ production on 
different forms of the unintegrated gluon distribution. Special attention
is given to the unintegrated gluon distributions obtained from BFKL evolution
equation which has been applied earlier in our previous papers~[15--18]. 
For studying $J/\Psi$ meson polarization properties we calculate 
the ${\bf p}_T^2$- and $Q^2$-dependences of the spin aligment parameter 
$\alpha$. 
 
The outline of this paper is as follows. In Section 2 we give the 
formulas for the inelastic $J/\Psi$ electroproduction differential 
cross section in the $k_T$-factorization QCD approach. In Section 3
we describe the unintegrated gluon distributions which we use in
our calculations. In Section 4 we give the formulas for 
$J/\Psi$ electroproduction off mass shell matrix elements in the CS model.
In Section 5 we present the results of our calculations. Finally, in
Section 6, we give some conclusions.

\bigskip

\section{Deep inelastic $J/\Psi$ electroproduction 
cross section in the $k_T$-factorization QCD approach} \indent 

In this section we calculate total and differential cross section for
inelastic $J/\Psi$ electroproduction via the photon-gluon fusion QCD
subprocess (Fig.~1) in the $k_T$-factorization QCD approach.
Let us define Sudakov variables of the process $e\,p\to e'\,J/\Psi\,X$:
\begin{equation}
p_{J/\Psi} = \alpha_1 p_e + \beta_1 p_p + p_{J/\Psi\,T},\qquad p_g = \alpha_2 p_e + \beta_2 p_p + p_{g\,T},\atop
\quad {\quad {q_1 = x_1 p_e + q_{1T},\qquad q_2 = x_2 p_p + q_{2T}}},
\end{equation}

\noindent where
\begin{equation}
p_{J/\Psi}^2 = m_{J/\Psi}^2,\quad p_g^2 = 0,\quad q_1^2 = q_{1T}^2 = - Q^2,\quad q_2^2 = q_{2T}^2,
\end{equation}

\noindent 
$p_{J/\Psi}$ and $p_g$ are 4-momenta of the $J/\Psi$ meson and 
final real gluon, $q_1$ and $q_2$ are 4-momenta of the initial virtual
photon and gluon, $p_{J/\Psi\,T}$, $p_{g\,T}$, $q_{1\,T}$, $q_{2\,T}$ are
transverse 4-momenta of these. In the center-of-mass frame of colliding
particles we can write:
\begin{equation}
p_e = (E,\,0,\,0,\,E),\qquad p_p = (E,\,0,\,0,\,-E),
\end{equation}

\noindent where
\begin{equation}
E = {\sqrt s\over 2},\qquad p_e^2 = p_p^2 = 0,\qquad (p_e\,\cdot\,p_p) = {s\over 2}.
\end{equation}

\noindent 
Sudakov variables are expressed as follows:
\begin{equation}
\displaystyle \alpha_1={m_{J/\Psi\,T}\over {\sqrt s}}\exp(y_{J/\Psi}),\qquad \alpha_2={{\bf p}_{g\,T}\over {\sqrt s}}\exp(y_g),\atop
\displaystyle \beta_1={m_{J/\Psi\,T}\over {\sqrt s}}\exp(-y_{J/\Psi}),\qquad \beta_2={{\bf p}_{g\,T}\over {\sqrt s}}\exp(-y_g),
\end{equation}

\noindent
where $m_{J/\Psi\,T}^2 = m_{J/\Psi}^2 + {\bf p}_{J/\Psi\,T}^2$, 
$m_{J/\Psi}$ is $J/\Psi$ meson mass, $y_{J/\Psi}$ and $y_g$ are 
c.m. rapidities of $J/\Psi$ meson and final gluon respectively.
\noindent 
The inelastic $J/\Psi$ electroproduction differential cross section has the 
following form:
\begin{equation}
\displaystyle d\sigma(e\,p\to e'\,J/\Psi\,X) = {dx_2 \over x_2}\,\Phi(x_2,\,{\bf q}_{2T}^2,\,\mu^2)\,{d\phi_2\over 2\pi}\,d{\bf q}_{2T}^2\,d\hat \sigma(e\,g^*\to e'\,J/\Psi\,g'),
\end{equation}

\noindent
where $\phi_2$ is initial BFKL gluon azimuthal angle, $\Phi(x_2,{\bf q}_{2T}^2,\mu^2)$ is an unintegrated gluon distribution in
the proton, and
\begin{equation}
d\hat \sigma(e\,g^*\to e'\,J/\Psi\,g') = \displaystyle {(2\pi)^4\over 2 x_2 s}\,\sum {|M|^2_{{\rm SHA}}(e\,g^*\to e'\,J/\Psi\,g')}\,\times \atop
\displaystyle \times {d^3p'_e\over (2\pi)^3\,2{p'}_e^0}\,{d^3p_{J/\Psi}\over (2\pi)^3\,2p_{J/\Psi}^0}\,{d^3p_g\over (2\pi)^3\,2p_g^0}\,\delta^{(4)}(p_e + q_2 - p_e' - p_{J/\Psi} - p_g),
\end{equation}

\noindent
where $\sum {|M|^2_{{\rm SHA}}(e\,g^*\to e'\,J/\Psi\,g')}$ is the off mass 
shell matrix element. In (9) $\sum$ indicates an averaging over 
initial particles polarization and a sum over polarization of final ones.

From (8) and (9) we obtain the following formula for the inelastic 
$J/\Psi$ electroproduction differential cross section in 
the $k_T$-factorization QCD approach:
\begin{equation}
\displaystyle d\sigma(e\,p\to e'\,J/\Psi\,X) = {1\over 128\pi^3}\, {\Phi(x_2,\,{\bf q}_{2T}^2,\,\mu^2)\over (x_2\,s)^2\,(1 - x_1)}\,{dz\over z\,(1 - z)}\,dy_{J/\Psi}\,\times \atop
\displaystyle \times\, \sum {|M|^2_{{\rm SHA}}(e\,g^*\to e'\,J/\Psi\,g')}\,d{\bf p}_{J/\Psi\,T}^2\,dQ^2\,d{\bf q}_{2T}^2\,{d\phi_1\over 2\pi}\,{d\phi_2\over 2\pi}\,{d\phi_{J/\Psi}\over 2\pi},
\end{equation}

\noindent where
\begin{equation}
x_1 = \alpha_1 + \alpha_2,\qquad x_2 = \beta_1 + \beta_2,\qquad {\bf q}_{1T} + {\bf q}_{2T} = {\bf p}_{J/\Psi\,T} + {\bf p}_{g\,T},
\end{equation}

\noindent
$\phi_1$ and $\phi_{J/\Psi}$ are initial virtual photon
and outgoing $J/\Psi$ meson azimuthal angles, 
$z = (p_{J/\Psi} \cdot p_p)/(q_1 \cdot p_p)$. If we take the limit
${\bf q}_{2T}^2\to 0$ and if we average (10) over the transverse directions
of the vector ${\bf q}_{2T}$, we obtain the formula for the differential cross 
section in the SPM:
\begin{equation}
\displaystyle d\sigma(e\,p\to e'\,J/\Psi\,X) = {1\over 128\pi^3}\, {1\over (x_2\,s)^2\,(1 - x_1)}\,{dz\over z\,(1 - z)}\, x_2G(x_2,\,\mu^2)\times \atop
\displaystyle \times \,\sum {|M|_{{\rm PM}}^2(e\,g\to e'\,J/\Psi\,g')}\,dy_{J/\Psi}\,d{\bf p}_{J/\Psi\,T}^2\,dQ^2\,{d\phi_1\over 2\pi}\,{d\phi_{J/\Psi}\over 2\pi},
\end{equation}

\noindent
where $\sum {|M|^2_{{\rm PM}}(e\,g\to e'\,J/\Psi\,g')}$ is the matrix element 
in the SPM. Here $\sum$ indicates an averaging over 
initial particles polarization and a sum over polarization of final ones.
We average over the transverse directions of the vector ${\bf q}_{2T}$ using the following
expression:
\begin{equation}
\displaystyle \int d{\bf q}_{2T}^2\,\int {\displaystyle {d\phi_2\over 2\pi}}\,\Phi(x_2,\,{\bf q}_{2T}^2,\,\mu^2)\,\sum {|M|_{{\rm SHA}}^2(e\,g^*\to e'\,J/\Psi\,g')} = \atop 
\displaystyle = x_2G(x_2,\,\mu^2)\,\sum {|M|_{{\rm PM}}^2(e\,g\to e'\,J/\Psi\,g')},
\end{equation}

\noindent
where 
\begin{equation}
\int\limits_0^{2\pi} {d\phi_2\over 2\pi}\,{q_{2T}^{\mu}\,q_{2T}^{\nu}\over {\bf q}_{2T}^2} = {1\over 2}\,g^{\mu\nu}.
\end{equation}

\bigskip

\section{Unintegrated gluon distributions} \indent 

Various parametrizations of the unintegrated gluon distribution used in
our calculations are discussed below. 

We used the results of a BFKL-like parametrization of unintegrated
gluon distribution (so called JB parametrization) according to the 
prescription given in~[44]. The proposed method lies upon a straightforward 
perturbative solution of the BFKL equation where the collinear gluon density 
$xG(x,\mu^2)$ from the standard GRV set~[45] is used as the boundary 
condition in the integral form (1). The unintegrated
gluon density is calculated as a convolution of collinear gluon distribution
$xG(x,\mu^2)$ with universal weight factors:
\begin{equation}
\Phi(x,{\bf q}_T^2,\mu^2) = \int\limits_x^1 \,\varphi(\eta,{\bf q}_T^2,\mu^2)\,{x\over \eta}\,G\left({x\over \eta},\,\mu^2\right)\,d\eta,
\end{equation}

\noindent
where
\begin{equation}
\displaystyle \varphi(\eta,{\bf q}_T^2,\mu^2) = \cases{\displaystyle {\bar \alpha_{S}\over \eta\, {\bf q}_T^2}J_0\left(2\sqrt{\mathstrut \bar\alpha_{S}\ln(1/\eta)\ln(\mu^2/{\bf q}_T^2)} \right),&if ${\bf q}_T^2\le \mu^2$,\cr
\displaystyle {\bar\alpha_{S}\over \eta\, {\bf q}_T^2}I_0\left(2\sqrt {\mathstrut \bar \alpha_{S}\ln (1/\eta) \ln ({\bf q}_T^2/\mu^2)}\right),&if ${\bf q}_T^2 > \mu^2$,\cr}
\end{equation}

\noindent
where $J_0$ and $I_0$ stand for Bessel function of real and imaginary arguments
respectively, and $\bar\alpha_{S}=3\alpha_{S}/\pi$. The parameter
$\bar\alpha_{S}$ is connected with the Pomeron trajectory intercept: 
$\Delta = 4\bar\alpha_{S}\ln 2$ in the LO and 
$\Delta = 4\bar\alpha_{S}\ln 2 - N\bar\alpha_{S}^2$ in the NLO
approximations, where  $N \sim 18$~[46, 47]. The latter value of $\Delta$
have dramatic consequences for high energy phenomenology, however, some
resummation procedures proposed in the last years lead to positive value of
$\Delta \sim 0.2-0.3$~[47, 48]. Therefore in our calculations with (17) we
used only the solution of LO BFKL equation and considered $\Delta$ as a free
parameter varying it from $0.166$ to $0.53$. Pomeron intercept parameter
$\Delta = 0.35$ was obtained from the description of $p_T$ spectrum of
$D^*$ meson electroproduction at HERA~[20]. We used this value of the 
parameter $\Delta$ in present paper.

Then we use another set (the KMS parametrization) [49] is obtained 
from a unified BFKL and DGLAP description of $F_2$ data and includes the 
so called consistency constraint [50]. The consistency constraint 
introduce a large correction to the LO BFKL equation: about 70~of the 
full NLO corrections to the BFKL exponent $\Delta$ are effectively
included in this constraint, as is declared in [51].

Finally, the third unintegrated gluon function used here is one proposed 
by Golec-Biernat and Wusthoff (GBW) which takes into account gluon saturation 
effects and has been applied earlier in analysis of the inclusive and 
diffractive $ep$-scattering data~[52, 53]. It has the following form:
\begin{equation}
\Phi(x,{\bf q}_T^2,\mu^2) = {3\sigma_0 \over 4\pi^2}\,{1\over \alpha_{S}}\,R_0^2(x)\,{\bf q}_T^2\,\exp(-R_0^2(x)\,{\bf q}_T^2),
\end{equation}

\noindent
where 
\begin{equation}
R_0^2(x) = {1 \over {\rm GeV}^2}\,{\left({x\over x_0}\right)}^{\lambda/2}.
\end{equation}

\noindent
and $\sigma_0 = 23.03\,{\rm mb}$, $\lambda = 0.288$, 
$x_0 = 3.04 \cdot 10^{-4}$.

In expression (1) the accounting of the contribution of the unintegrated
gluon distribution to integral at low ${\bf q}_T^2$ region 
$0 < {\bf q}_T^2 < Q_0^2$ was done by the collinear gluon density 
$xG(x,Q_0^2)$, where $Q_0^2 = 1\,{\rm GeV}^2$ for all gluon distributions.

\bigskip

\section{Off mass shell matrix element for $e\,g^*\to e'\,J/\Psi\,g'$ process} \indent 

In this section we give the formulas for $J/\Psi$ electroproduction off mass 
shell matrix elements in the CS model. There are six Feynman diagrams 
(Fig.~2) which describe partonic process
$\gamma\,g^*\to J/\Psi\,g'$ at the leading order in $\alpha_{S}$ and 
$\alpha$. In the framework of the CS model and nonrelativistic approximation
the production of $J/\Psi$ meson is considered as the production of a 
quark-antiquark system in the colour singlet state with orbital momentum
$L = 0$ and spin momentum $S = 1$. The binding energy and relative
momentum of quarks in the $J/\Psi$ are neglected. Such a way 
$m_{J/\Psi} = 2\,m_c$, where $m_c$ is charm mass. Taking into account
the formalism of the projection operator~[41] the amplitude of the 
process $\gamma\,g^*\to J/\Psi\,g'$ may be obtained from the amplitude
of the process $\gamma\,g^*\to c\bar c\,g'$ after replacement:
\begin{equation}
v(p_{\bar c})\,\bar u(p_c) \to \hat J(p_{J/\Psi}) = {\Psi(0)\over 2\sqrt {\mathstrut m_{J/\Psi}}}\,\hat \epsilon(p_{J/\Psi})\,(\hat p_{J/\Psi} + m_{J/\Psi})\,{1\over \sqrt {\mathstrut 3}},
\end{equation}

\noindent
where $\hat \epsilon(p_{J/\Psi}) = \epsilon_{\mu}(p_{J/\Psi})\,\gamma^{\mu}$, 
$\epsilon(p_{J/\Psi})$ is a 4-vector of the $J/\Psi$ polarization, 
$1/\sqrt 3$ is the color factor, $\Psi(0)$ is the nonrelativistic meson
wave function at the origin. In according with the standard Feynman rules
for QCD diagrams we have:
\begin{equation}
\displaystyle M = e_c\,g^2\,\epsilon_{\mu}(q_1)\,\epsilon_{\sigma}(q_2)\,\epsilon_{\rho}(p_g)\,\times \atop 
\displaystyle \times Sp \left[\hat J(p_{J/\Psi})\,\gamma^{\mu}\,{\hat p_{c} - \hat q_1 + m_c\over (p_c - q_1)^2 - m_c^2}\,\gamma^{\sigma}\,{ - \hat p_c - \hat p_g + m_c\over (- p_c - p_g)^2 - m_c^2}\,\gamma^{\rho}\right]
\end{equation}

\noindent
+ 5 permutations of all gauge bosons. Here $\epsilon_{\mu}(q_1)$, 
$\varepsilon_{\mu}(q_2)$ are polarization vectors of the initial photon and 
gluon respectively, $\epsilon_{\mu}(p_g)$ is a 4-vector of the final gluon 
polarization. The summation on the $J/\Psi$ meson and final gluon 
polarizations is carried out by covariant formulas:
\begin{equation}
\sum {\epsilon^{\mu}(p_{J/\Psi})\epsilon^{*\,\nu}(p_{J/\Psi})} = - g^{\mu\nu} + {p_{J/\Psi}^{\mu}p_{J/\Psi}^{\nu}\over m_{J/\Psi}^2},
\end{equation}
\begin{equation}
\sum {\epsilon^{\mu}(p_g)\epsilon^{*\,\nu}(p_g)} = - g^{\mu\nu},
\end{equation}

\noindent
The initial BFKL gluon polarization tensor is taken in form (2). We
substitute here the full lepton tensor for the photon polarization tensor
(including also the photon propagator factor and photon-lepton coupling):
\begin{equation}
\sum {\epsilon^{\mu}(q_1)\epsilon^{*\,\nu}(q_1)} = 2\,{e^2\over Q^2}\,\left( - g^{\mu\nu} + {4 p_{e}^{\mu}p_{e}^{\nu}\over Q^2}\right).
\end{equation}

\noindent
For studing $J/\Psi$ polarized production we introduce the 4-vector
of the longitudinal polarization $\epsilon_L^{\mu}(p_{J/\Psi})$ as follows~[54]:
\begin{equation}
\epsilon^{\mu}_L(p_{J/\Psi}) = {(p_{J/\Psi}\cdot p_p)\over \sqrt {\mathstrut {(p_{J/\Psi}\cdot p_p)^2 - m_{J/\Psi}^2}\,s}}\left({p_{J/\Psi}^{\mu} \over m_{J/\Psi}} - {m_{J/\Psi}p_p^{\nu}\over (p_{J/\Psi}\cdot p_p)}\right).
\end{equation}

\noindent
In the limit $s \gg m_{J/\Psi}^2$ the polarization 4-vector satisfies usual
conditions: $(\epsilon_L \cdot \epsilon_L) = - 1$ and $(\epsilon_L \cdot p_{J/\Psi}) = 0$.
The calculation of $\sum{|M|_{{\rm SHA}}^2}(e\,g^*\to e'\,J/\Psi\,g')$
and $\sum {|M|_{{\rm PM}}^2(e\,g\to e'\,J/\Psi\,g')}$ was done 
analitically by REDUCE system.

\bigskip
\renewcommand{\thefootnote}{3)}
\section{Numerical results and discussion} \indent 

In this section we present the theoretical results in comparison with 
available experimental data taken by the H1~[42] and ZEUS~[43] collaborations 
at HERA. There are three parameters which determine the common 
normalization factor of the cross section under consideration: $J/\Psi$ 
meson wave function at the origin $\Psi(0)$, charmed quark mass $m_c$ and 
factorization scale $\mu$. The value of the $J/\Psi$ meson wave function 
at the origin may be calculated in a potential model or obtained from 
the well known experimental 
decay width $\Gamma(J/\Psi \to \mu^{+}\,\mu^{-})$. In our calculation we used 
the following choice $|\Psi(0)|^2 = 0.0876\,{\rm GeV}^3$ as in~[55].
Concerning a charmed quark mass, the situation is not clear: 
on the one hand, in the nonrelativistic approximation one has 
$m_c = m_{J/\Psi}/2 = 1.55\,{\rm GeV}$, but on the other hand there are 
many examples when smaller value of a charm mass is used, 
for example, $m_c = 1.4\,{\rm GeV}$~[34, 56, 57]. 
In the present paper we used both of charm mass values. Also the most 
significant theoretical uncertanties refer to the choice of the factorization 
scale $\mu_F$ and renormalization one $\mu_R$. One of them is related to
the evolution of the gluon distributions $\Phi(x,{\bf q}_T^2,\mu_F^2)$, the
other is responsible for strong coupling constant $\alpha_{S}(\mu_R^2)$.
As is often done in literature for simplicity we set them equal: $\mu_F = \mu_R = \mu$.
In the present paper we used the following choice $\mu^2 = {\bf q}_{2T}^2$ 
as in~[58].\footnote{The our preliminary results at~[59] were obtained with
$\mu^2 = m_{J/\Psi}^2 + {\bf p}_{J/\Psi\,T}^2$.}

The calculations of the inelastic $J/\Psi$ production cross section in the
$k_T$-factorization approach have been made according to (10) and in the 
SPM --- according to (12). The ${\bf p}_{J/\Psi\,T}^2$ integration limits
are taken as given by inelastic $J/\Psi$ process requirements:
$ 1\,{\rm GeV}^2 \le {\bf p}_{J/\Psi\,T}^2 \le s/4 - m_{J/\Psi}^2$.
According to (1) the ${\bf q}_{2T}^2$ integration region was divided into 
two parts: $0 < {\bf q}_{2T}^2 \le Q_0^2$ and ${\bf q}_{2T}^2 \ge Q_0^2$. The
calculations are made according to (10) for ${\bf q}_{2T}^2 \ge Q_0^2$ and
for $0 < {\bf q}_{2T}^2 \le Q_0^2$ we use $\sum {|M|_{{\rm PM}}^2(e\,g\to e'\,J/\Psi\,g')}$
instead of $\sum {|M|_{{\rm SHA}}^2(e\,g^*\to e'\,J/\Psi\,g')}$ (see also (12)).
The choice of the critical value of parameter $Q_0^2 = 1\,{\rm GeV}^2$ is
determined by the requirement that the value of $\alpha_{S}({\bf q}_{2T}^2)$
be small in the region ${\bf q}_{2T}^2 \ge 1\,{\rm GeV}^2$, where in fact
$\alpha_{S}({\bf q}_{2T}^2) \le 0.26$.

The $Q^2$, $y_{J/\Psi}$ and $z$ integration limits are taken as given by
kinematical conditions of H1 experimental data~[42]. The first kinematical 
region is $2\,{\rm GeV}^2 \le Q^2 \le 80\,{\rm GeV}^2$, 
$40 \,{\rm GeV} \le W \le 180 \,{\rm GeV}$, $z > 0.2$, $M_X \ge 10\,{\rm GeV}$ 
and the second one is ${\bf p}_{J/\Psi\,T}^2 \ge 4\,{\rm GeV}^2$,
$4\,{\rm GeV}^2 \le Q^2 \le 80\,{\rm GeV}^2$, 
$40 \,{\rm GeV} \le W \le 180 \,{\rm GeV}$, $z > 0.2$, $M_X \ge 10\,{\rm GeV}$.

The results of our calculations are shown in Fig.~3---8. Fig.~3 show the
$Q^2$-, ${\bf p}_{J/\Psi\,T}^2$-, $z$-, $y_{J/\Psi}^*$- 
and $W$-distributions of the inelastic $J/\Psi$ meson production obtained in
the first kinematical region at $\sqrt s = 314\,{\rm GeV}$, 
$m_c = 1.55\,{\rm GeV}$ and $\Lambda_{{\rm QCD}} = 250\,{\rm MeV}$. 
Curve {\sl 1} corresponds to the SPM calculations at 
the leading order approximation with GRV gluon density, curves {\sl 2}, 
{\sl 3} and {\sl 4} correspond to the $k_T$-factorization  
results with JB, KMS and GBW unintegrated gluon distributions at 
$Q_0^2 = 1\, {\rm GeV}^2$. One can see that results
obtained in the CS model with $k_T$-factorization (curves {\sl 2 --- 4}) 
agree in shape but underestimate the H1 experimental data. 
The SPM calculation (curve {\sl 1}) are lower the data by a factor 2.

Fig.~4 show the $z$-, $y_{J/\Psi}^*$- and $W$-distributions of the 
inelastic $J/\Psi$ meson production obtained in the second 
kinematical region at $\sqrt s = 314\,{\rm GeV}$, $m_c = 1.55\,{\rm GeV}$ and 
$\Lambda_{{\rm QCD}} = 250\,{\rm MeV}$. 
Curves {\sl 1 --- 4} are the same as in Fig.~3. One can see that in 
this kinematical region the $k_T$-factorization 
approach (curves {\sl 2 --- 4}) describe the data better than in the
first one (without cut ${\bf p}_{J/\Psi\,T}^2 \ge 4\,{\rm GeV}^2$).

Fig.~5,~6 shows the $Q^2$-, ${\bf p}_{J/\Psi\,T}^2$-, $z$-, $y_{J/\Psi}^*$- 
and $W$-distributions of the inelastic $J/\Psi$ meson production 
at $\sqrt s = 314\,{\rm GeV}$, $m_c = 1.4\,{\rm GeV}$ and 
$\Lambda_{{\rm QCD}} = 250\,{\rm MeV}$ obtained in
the first kinematical region (Fig.~5) and in the second one (Fig.~6).
Curves {\sl 1 --- 4} are the same as in Fig.~3. 
One can see that a shift the charm quark mass down to $m_c = 1.4\,{\rm GeV}$ 
increases the cross section by a factor 1.5, and results obtained in 
the CS model with $k_T$-factorization (curves {\sl 2 --- 4}) agree 
with H1 experimental data.

The $z$-distributions are 
described at $z > 0.4$ only. It is because other $J/\Psi$ 
production mechanisms (such as resolved photon and/or colour octet 
contributions) may be impotant at $z < 0.4$~[60]. 

It is notable that the saturation unintegrated gluon distribution (GBW 
parametrization) does not contradict the HERA experimental data.

Fig.~3 --- 6 show that $k_T$-factorization results with $m_c = 1.4\,{\rm GeV}$
(in contrast with SPM ones) for inelastic $J/\Psi$ electroproduction
at HERA reasonably agree with H1 experimental data. Hovewer, 
at $m_c = 1.55\,{\rm GeV}$ some additional transition mechanism 
from $c\bar c$-pair to the $J/\Psi$ mesons (such as CO model) need for full 
description of these data.

\renewcommand{\thefootnote}{4)}
We analyzed specially the influence of charm quark mass, $m_c$, on the our
theoretical results. We found that the main effect of change of the charm quark mass
connects with final phase space of $J/\Psi$ meson. In the subprocess matrix elements 
this effect is neglectable. However the value of $m_c = 1.4\,{\rm GeV}$ corresponds to 
the unphysical phase space of $J/\Psi$ state.\footnote{We thank S.~Baranov for 
the suggestion to study this problem.}

For studying $J/\Psi$ meson polarization properties we calculate 
the ${\bf p}_T^2$- and $Q^2$-dependences of the spin aligment parameter 
$\alpha$~[19, 61] in the $k_T$-factorization approach and SPM:
\begin{equation}
\alpha (\omega) = {d\sigma/d\omega - 3\,d\sigma_L/d\omega\over d\sigma/d\omega + d\sigma_L/d\omega},
\end{equation}
\noindent
where $\sigma_L$ is the production cross section for the longitudinally
polarized $J/\Psi$ mesons, $\omega = {\bf p}_{J/\Psi\,T}^2,\,Q^2$.
The parameter $\alpha$ controls the angular distribution for leptons
in the decay $J/\Psi \to \mu^{+}\,\mu^{-}$ (in the $J/\Psi$ meson 
rest frame):
\begin{equation}
{d\Gamma(J/\Psi \to \mu^{+}\,\mu^{-})\over d\cos\theta} \sim 1 + \alpha\,\cos^2\theta.
\end{equation}
\noindent

Fig.~7 shows the parameter $\alpha({\bf p}_{J/\Psi\,T})$, which is
calculated in the region $0.4 < z < 0.9$ (a) and in the
region $0.4 < z < 1$ (b) at $\sqrt s = 314\,{\rm GeV}$, 
$m_c = 1.4\,{\rm GeV}$ and $\Lambda_{{\rm QCD}} = 250\,{\rm MeV}$
in comparison with preliminary experimental data taken by the ZEUS~[43] 
collaboration at HERA. Curve {\sl 1} corresponds to the SPM calculations at 
the leading order approximation with GRV gluon density, curve {\sl 3} 
corresponds to the $k_T$-factorization QCD calculations with KMS unintegrated 
gluon distribution at $Q_0^2 = 1\, {\rm GeV}^2$.

We note that it is impossible to make of exact conclusions about a BFKL gluon 
contribution to the polarized $J/\Psi$ 
production cross section because of large incertainties in the experimental data and 
large additional contribution from initial 
longitudinal polarization of virtual photons.
However at low $Q^2 < 1\,{\rm GeV}^2$ such contributions are neglected.
This fact should result in observable spin effects of final $J/\Psi$ mesons.
As example, we have performed calculations for the spin parameter $\alpha$ as a 
function ${\bf p}_{J/\Psi\,T}^2$ at fixed values of $Q^2$ for
$40 \,{\rm GeV} \le W \le 180 \,{\rm GeV}$, 
$z > 0.2$, $M_X \ge 10\,{\rm GeV}$ at $\sqrt s = 314\,{\rm GeV}$, 
$m_c = 1.4\,{\rm GeV}$ and $\Lambda_{{\rm QCD}} = 250\,{\rm MeV}$.

The results of our 
calculations at fixed values of $Q^2 = 0.1\,{\rm GeV}^2$, 
$Q^2 = 1\,{\rm GeV}^2$, $Q^2 = 5\,{\rm GeV}^2$ and $Q^2 = 10\,{\rm GeV}^2$
are shown in Fig.~8. Curves {\sl 1} and {\sl 3} are the same as in Fig.~7. 
We have large difference between predictions 
of the leading order of SPM and the $k_T$-factorization approach at low 
$Q^2 < 1\,{\rm GeV}^2$, as it is seen in Fig.~8.
Therefore if account of the NLO corrections (which is still unknown) will 
not change predictions of the LO SPM for $\alpha({\bf p}_{J/\Psi\,T}^2)$ 
at low $Q^2 < 1\,{\rm GeV}^2$, experimental measurement of 
polarization properties of the $J/\Psi$ mesons will be an 
additional test of BFKL gluon dynamics.

\bigskip

\section{Conclusions} \indent 

In this paper we considered $J/\Psi$ meson production in $ep$ deep inelastic 
scattering in the colour singlet model using the standart parton model at the 
leading order in $\alpha_{S}$ and the $k_T$-factorization QCD approach.
We investigated the $z$-, $Q^2$-, ${\bf p}_T^2$-, $y^*$-, $W$-dependences of
inelastic $J/\Psi$ production on different forms of the unintegrated 
gluon distribution. The ${\bf p}_T^2$- and $Q^2$-dependences of the 
spin aligment parameter $\alpha$ presented also. We compared the 
theoretical results with available experimental data taken by the H1 and ZEUS 
collaborations at HERA. We have found that the $k_T$-factorization 
results with JB, KMS and GBW unintegrated gluon distributions 
reasonably agree with H1 experimental data only at $m_c = 1.4\,{\rm GeV}$, 
$|\Psi(0)|^2 = 0.0876\,{\rm GeV}^3$ and 
$\Lambda_{{\rm QCD}} = 250\,{\rm MeV}$. At $m_c = 1.55\,{\rm GeV}$ some 
additional transition mechanism 
from $c\bar c$-pair to the $J/\Psi$ mesons (such as CO model) need for full 
description of these data.
The saturation unintegrated gluon distribution (GBW 
parametrization) does not contradict the existing experimental data.
The experimental study of a polarization of $J/\Psi$ meson  
at low $Q^2 < 1\,{\rm GeV}^2$ should be additianal test of BFKL gluon dynamics.

The authors would like to thank S.~Baranov for encouraging interest and 
useful discussions. A.L. thanks also V.~Saleev for the help on the 
initial stage of work. The study was supported in part by RFBR grant
02--02--17513.

\newpage
\thispagestyle{empty}

\noindent
{\bf Fig.~1.} Diagram for $e\,p\to e'\,J/\Psi\,X$ process.

\vspace{1cm}
\noindent
{\bf Fig.~2.} Feynmans diagram used for description partonic process 
$\gamma\,g\to J/\Psi\,g'$ process.

\vspace{1cm}
\noindent
{\bf Fig.~3.} The $Q^2$-, ${\bf p}_{J/\Psi\,T}^2$-, $z$-, $y_{J/\Psi}^*$- 
and $W$-distributions of the inelastic $J/\Psi$ production obtained in
the kinematical region $2\,{\rm GeV}^2 \le Q^2 \le 80\,{\rm GeV}^2$, 
$40 \,{\rm GeV} \le W \le 180 \,{\rm GeV}$, $z > 0.2$, $M_X \ge 10\,{\rm GeV}$
at $\sqrt s = 314\,{\rm GeV}$, 
$m_c = 1.55\,{\rm GeV}$ and $\Lambda_{{\rm QCD}} = 250\,{\rm MeV}$. 
Curve {\sl 1} corresponds to the SPM calculations at 
the leading order approximation with GRV gluon density, curves {\sl 2}, 
{\sl 3} and {\sl 4} correspond to the $k_T$-factorization QCD calculations 
with JB, KMS and GBW unintegrated gluon distribution at 
$Q_0^2 = 1\, {\rm GeV}^2$.

\vspace{1cm}
\noindent
{\bf Fig.~4.} The $z$-, $y_{J/\Psi}^*$- and $W$-distributions of the 
inelastic $J/\Psi$ production obtained in
the kinematical region ${\bf p}_{J/\Psi\,T}^2 \ge 4\,{\rm GeV}^2$,
$4\,{\rm GeV}^2 \le Q^2 \le 80\,{\rm GeV}^2$, 
$40 \,{\rm GeV} \le W \le 180 \,{\rm GeV}$, $z > 0.2$, $M_X \ge 10\,{\rm GeV}$
at $\sqrt s = 314\,{\rm GeV}$, 
$m_c = 1.55\,{\rm GeV}$ and $\Lambda_{{\rm QCD}} = 250\,{\rm MeV}$. 
Curves {\sl 1 --- 4} are the same as in Fig.~3.

\vspace{1cm}
\noindent
{\bf Fig.~5.} The $Q^2$-, ${\bf p}_{J/\Psi\,T}^2$-, $z$-, $y_{J/\Psi}^*$- 
and $W$-distributions of the inelastic $J/\Psi$ production obtained in
the kinematical region $2\,{\rm GeV}^2 \le Q^2 \le 80\,{\rm GeV}^2$, 
$40 \,{\rm GeV} \le W \le 180 \,{\rm GeV}$, $z > 0.2$, $M_X \ge 10\,{\rm GeV}$
at $\sqrt s = 314\,{\rm GeV}$, 
$m_c = 1.4\,{\rm GeV}$ and $\Lambda_{{\rm QCD}} = 250\,{\rm MeV}$. 
Curves {\sl 1 --- 4} are the same as in Fig.~3.

\vspace{1cm}
\noindent
{\bf Fig.~6.} The $z$-, $y_{J/\Psi}^*$- and $W$-distributions of the 
inelastic $J/\Psi$ production obtained in
the kinematical region ${\bf p}_{J/\Psi\,T}^2 \ge 4\,{\rm GeV}^2$,
$4\,{\rm GeV}^2 \le Q^2 \le 80\,{\rm GeV}^2$, 
$40 \,{\rm GeV} \le W \le 180 \,{\rm GeV}$, $z > 0.2$, $M_X \ge 10\,{\rm GeV}$
at $\sqrt s = 314\,{\rm GeV}$, 
$m_c = 1.4\,{\rm GeV}$ and $\Lambda_{{\rm QCD}} = 250\,{\rm MeV}$. 
Curves {\sl 1 --- 4} are the same as in Fig.~3.

\vspace{1cm}
\noindent
{\bf Fig.~7.} The parameter $\alpha({\bf p}_{J/\Psi\,T})$, which is
calculated in the region $0.4 < z < 0.9$ (a) and in the
region $0.4 < z < 1$ (b) at $\sqrt s = 314\,{\rm GeV}$, 
$m_c = 1.4\,{\rm GeV}$ and $\Lambda_{{\rm QCD}} = 250\,{\rm MeV}$. 
Curve {\sl 1} corresponds to the SPM calculations at the leading order approximation 
with GRV gluon density, curve {\sl 3} corresponds to the $k_T$-factorization QCD 
calculations with KMS unintegrated gluon distribution at $Q_0^2 = 1\, {\rm GeV}^2$.

\vspace{1cm}
\noindent
{\bf Fig.~8.} The spin aligment parameter $\alpha({\bf p}_{J/\Psi\,T}^2)$
at fixed values of $Q^2$ for
$40 \,{\rm GeV} \le W \le 180 \,{\rm GeV}$, 
$z > 0.2$, $M_X \ge 10\,{\rm GeV}$ at $\sqrt s = 314\,{\rm GeV}$, 
$m_c = 1.4\,{\rm GeV}$ and $\Lambda_{{\rm QCD}} = 250\,{\rm MeV}$.
Curves {\sl 1} and {\sl 3} are the same as in Fig.~7.

\end{large}
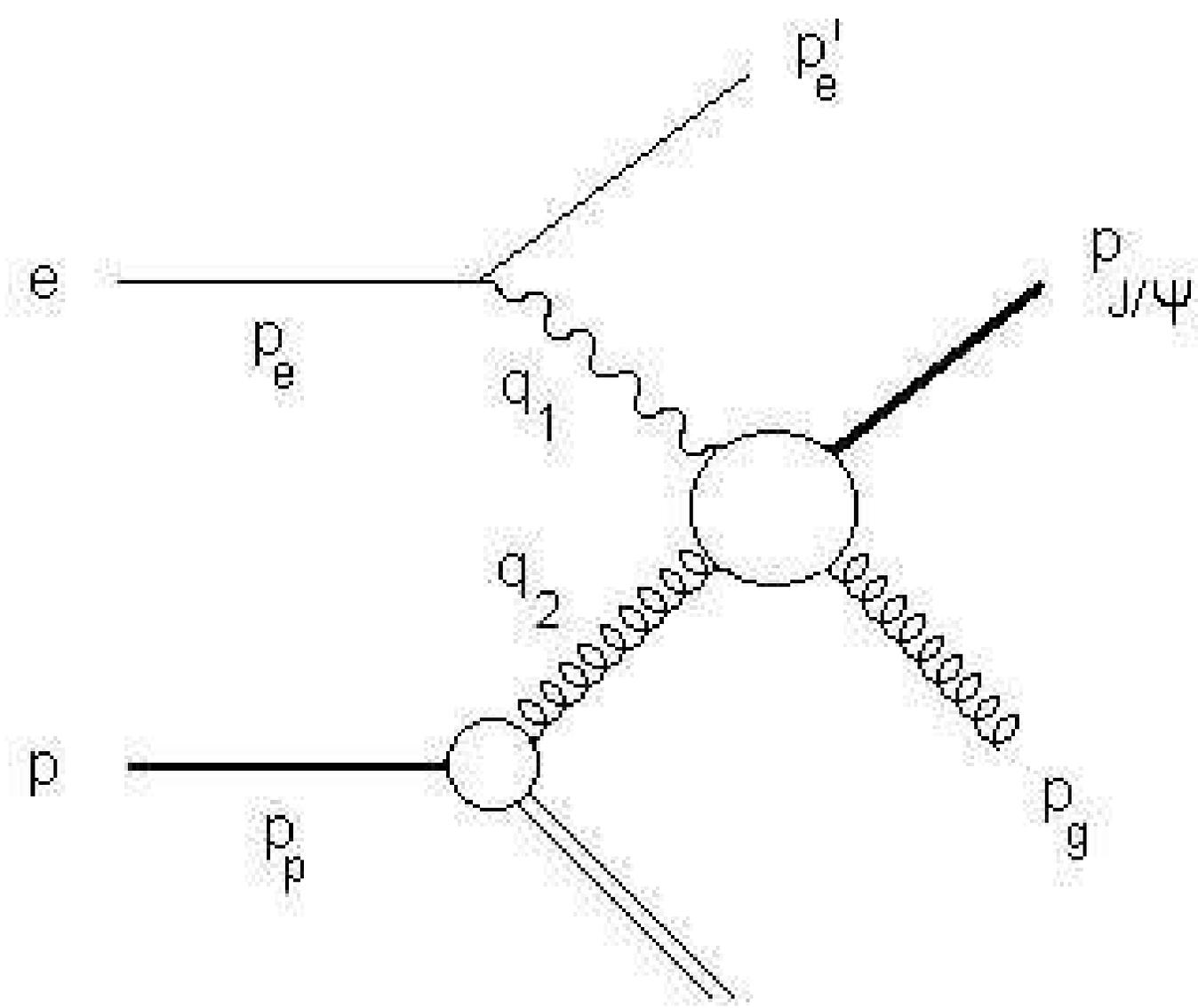
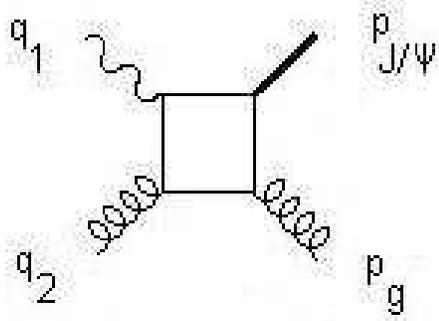
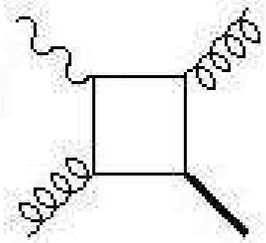
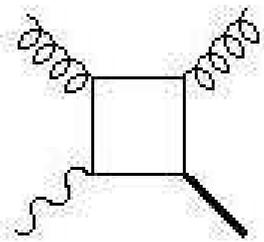
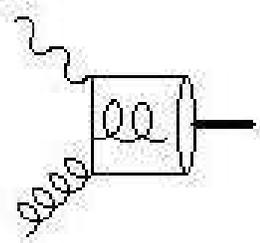
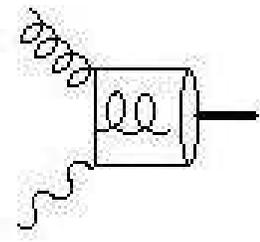
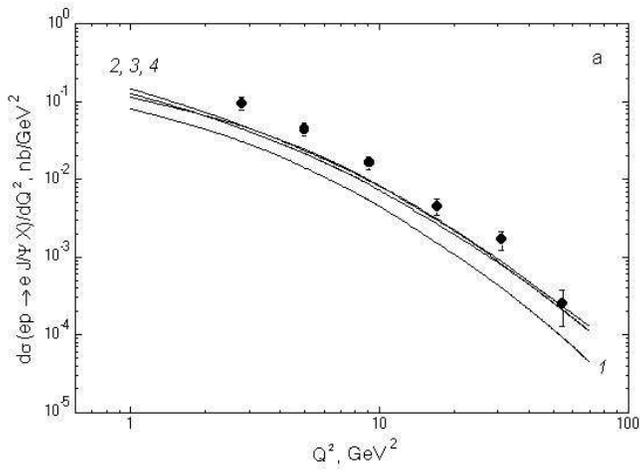
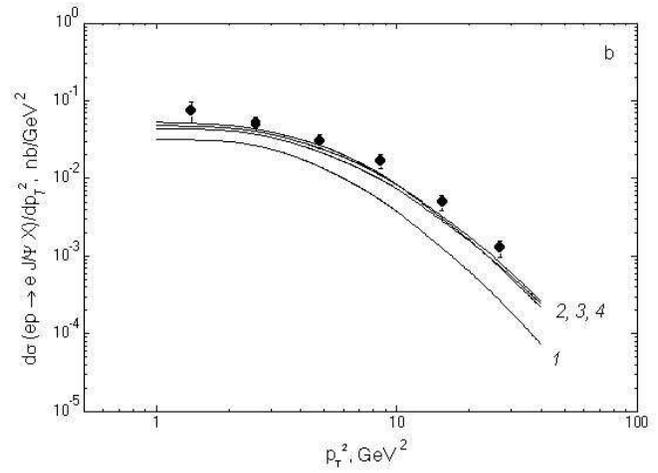
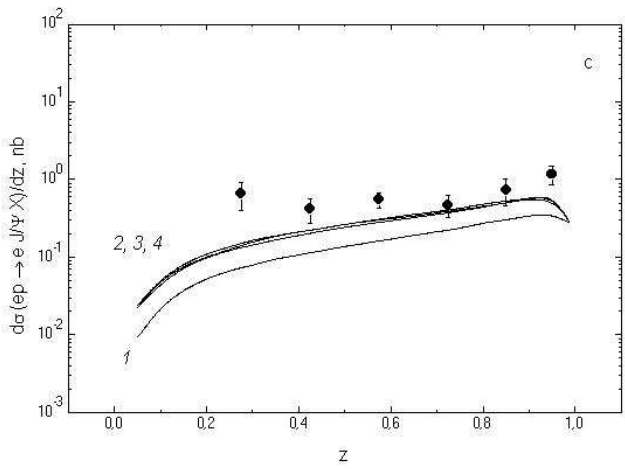
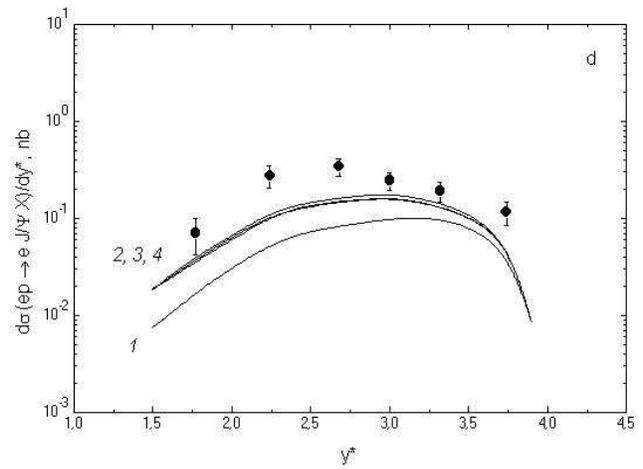
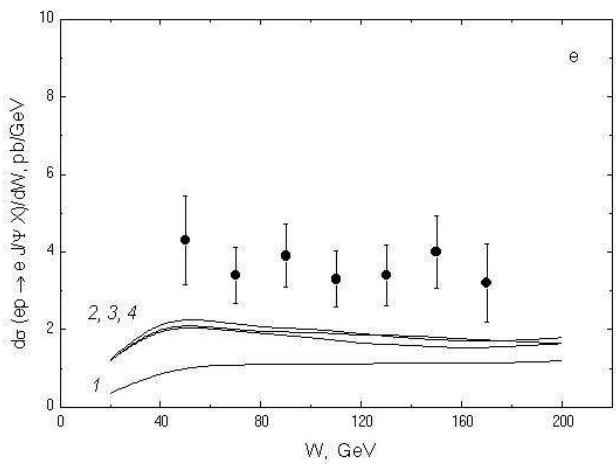
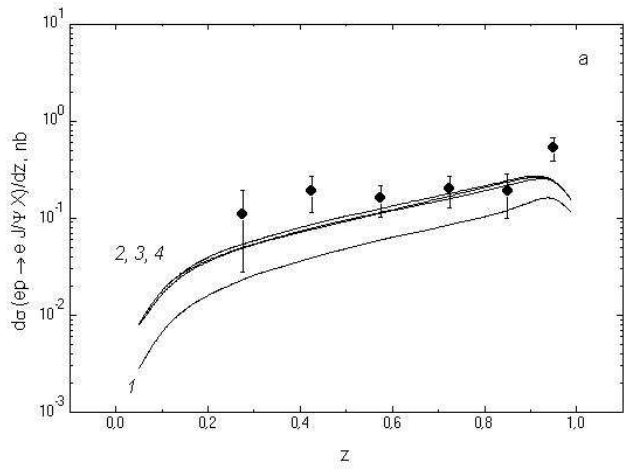
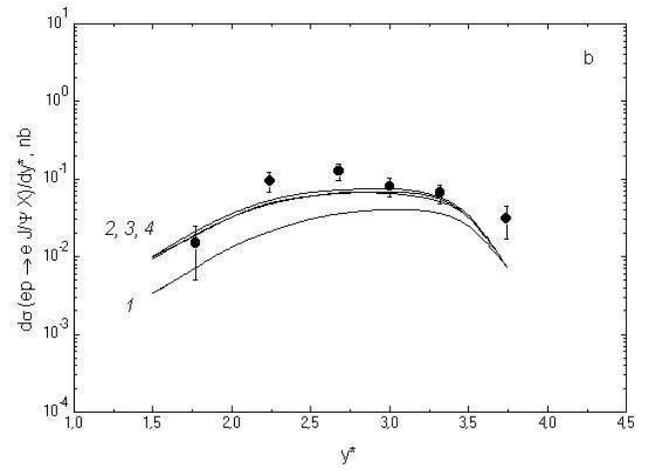
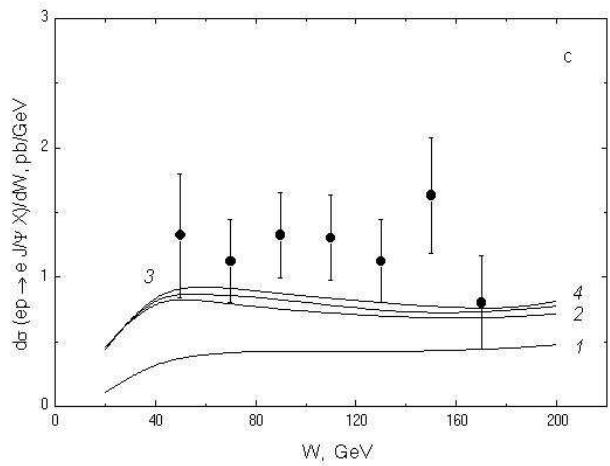
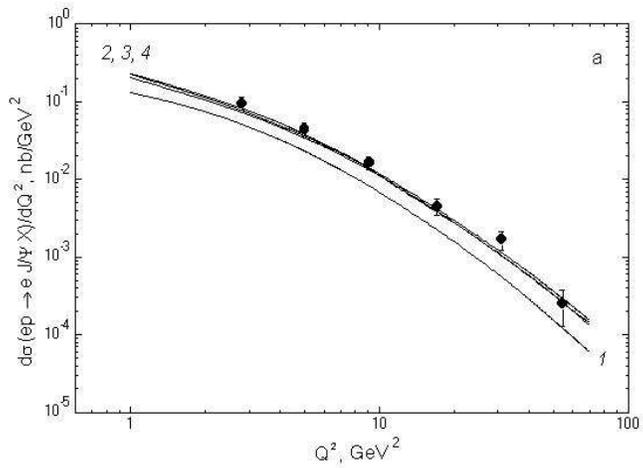
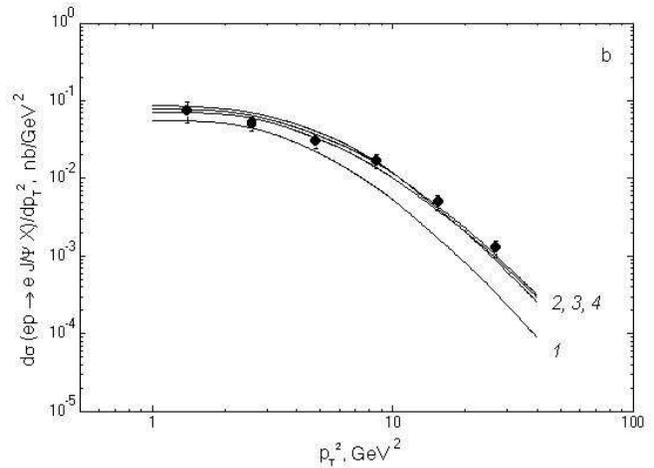
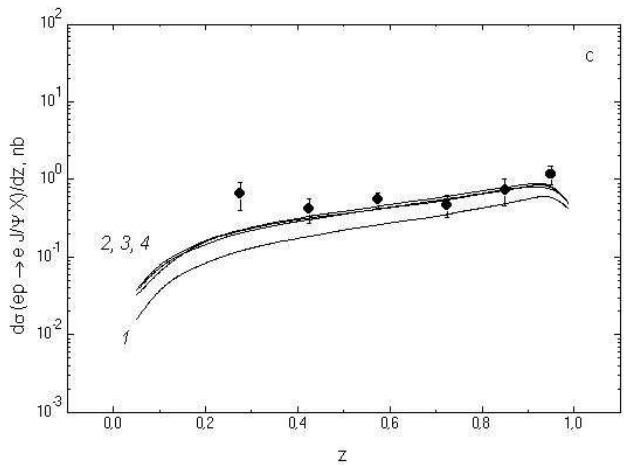
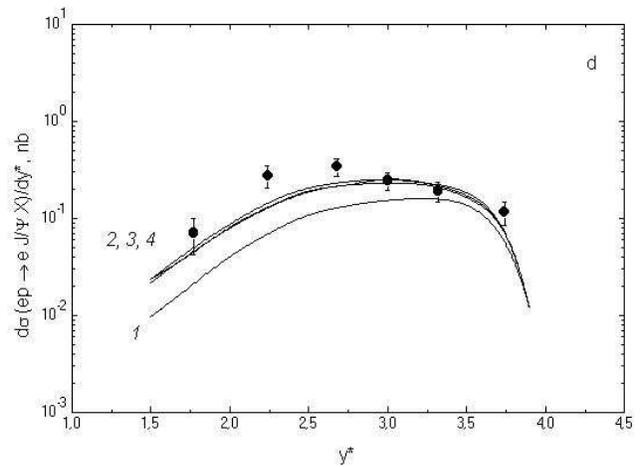
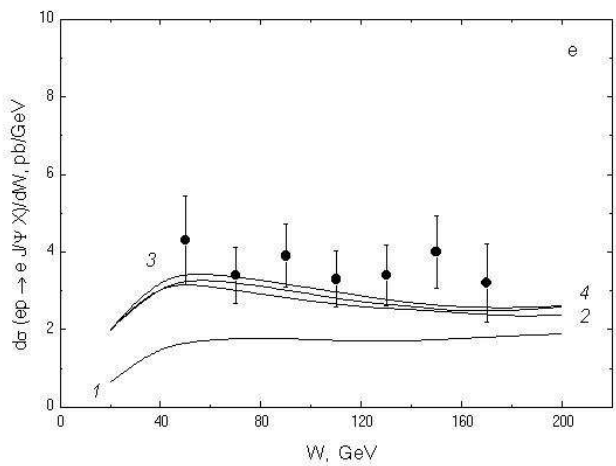
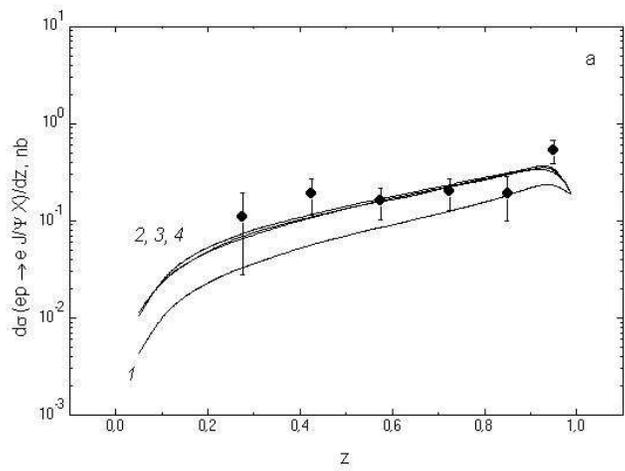
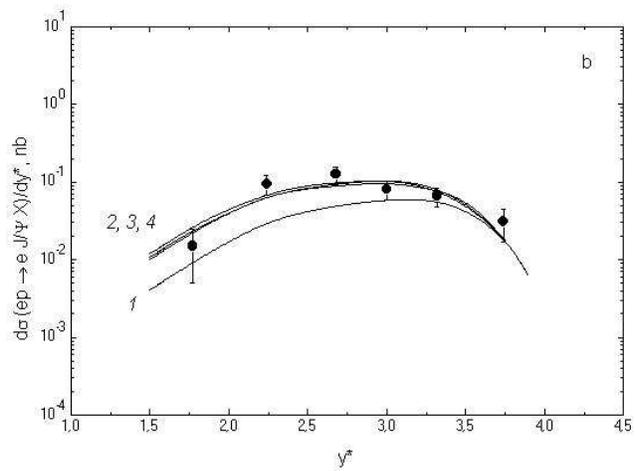
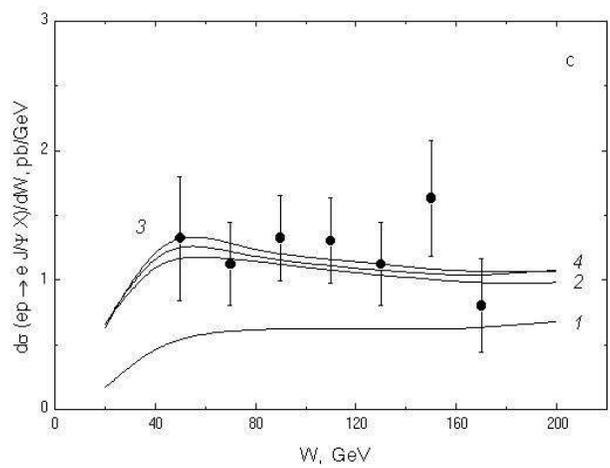
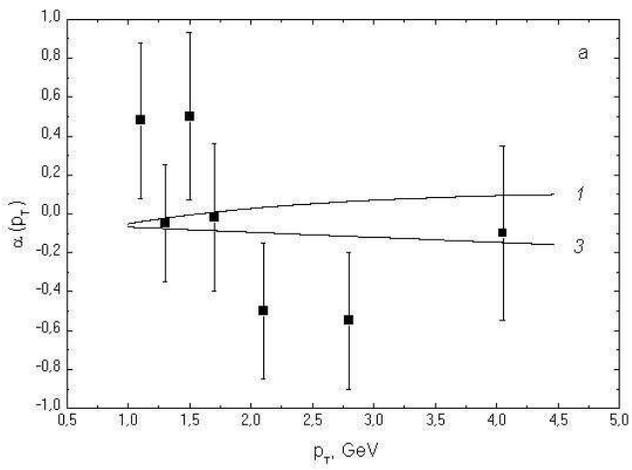
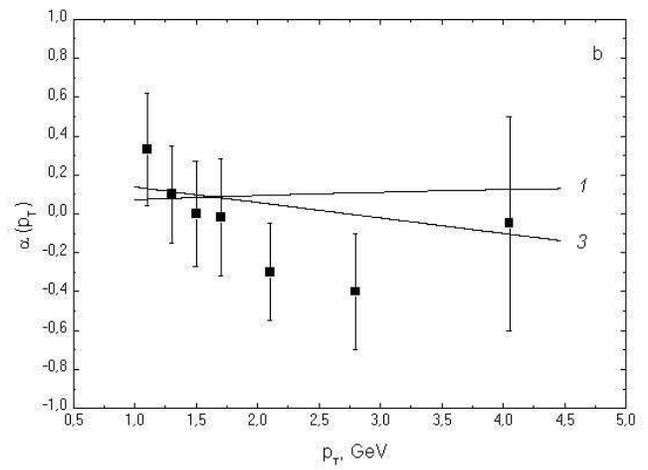
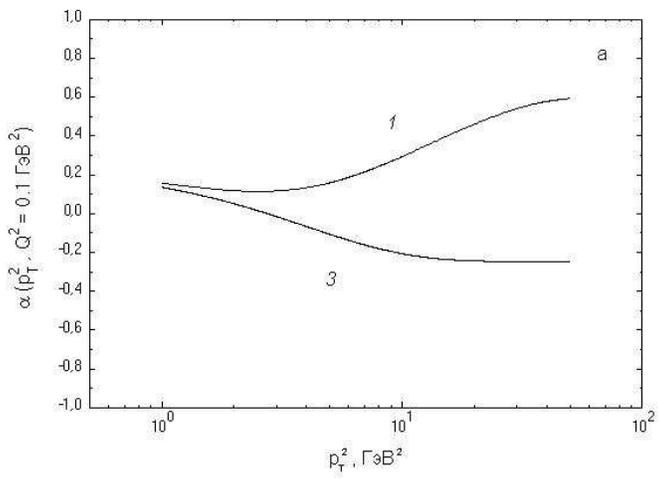
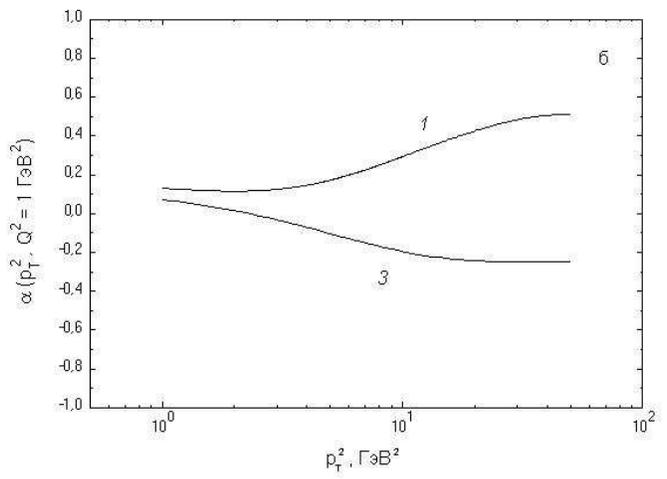
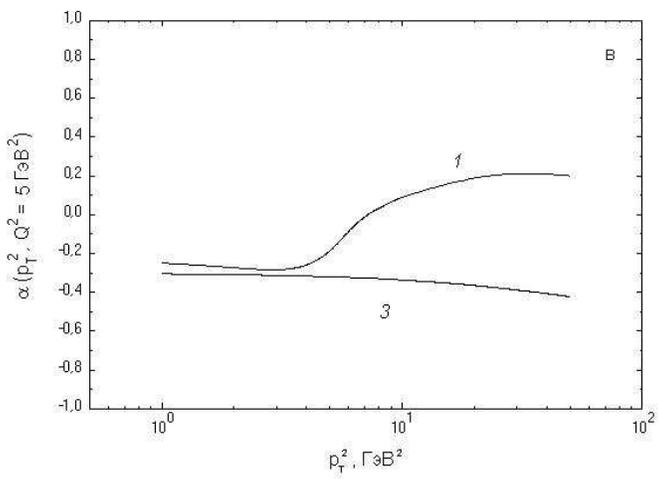
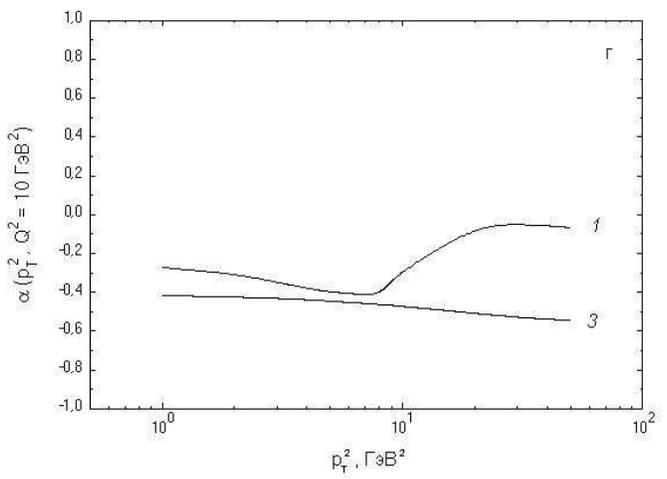
\end{document}